\documentclass[12pt]{iopart}

\usepackage{graphicx}
\usepackage{amssymb}
\usepackage{graphicx,epstopdf,epsfig}
\begin{document}

\title{Dynamics of levitated nanospheres: towards the strong coupling regime}

\author{T. S. Monteiro}
\address{Department of Physics and Astronomy,
University College London, Gower Street, London WC1E 6BT, United Kingdom}

\author{J. Millen}
\address{Department of Physics and Astronomy,
University College London, Gower Street, London WC1E 6BT, United Kingdom}

\author{G. A. T. Pender}
\address{Department of Physics and Astronomy,
University College London, Gower Street, London WC1E 6BT, United Kingdom}
\author{Florian Marquardt}
\address{Institut for Theoretical Physics, Universit\"at Erlangen-N\"urnberg,
Staudtstra\ss e 7, 91058 Erlangen Germany}
\address{Max Planck Institute for the Science of Light,
G\"unther-Scharowsky-Stra\ss e 1/Bau 24, 91058 Erlangen Germany}

\author{D. Chang}
\address{ICFO - Institut de Ciencies Fotoniques Av. Carl Friedrich Gauss,
308860 Castelldefels, (Barcelona), Spain }

\author{P. F. Barker}
\address{Department of Physics and Astronomy,
University College London, Gower Street, London WC1E 6BT, United Kingdom}

\newpage

\begin{abstract}
{ The use of levitated nanospheres represents a new paradigm for the
optomechanical cooling of a small mechanical oscillator, with the prospect of
realising quantum oscillators with unprecedentedly high quality factors.
We investigate the dynamics of this system, especially in the so-called
self-trapping regimes,
where one or more optical fields simultaneously trap and cool the mechanical
oscillator. The determining characteristic of
this regime is that both the mechanical frequency $\omega_M$ and
single-photon optomechanical coupling strength parameters
$g$ are a function of the optical field intensities, in contrast to usual
set-ups where $\omega_M$ and $g$
are constant for the given system.
We also measure the characteristic transverse and axial trapping frequencies of
different
sized silica nanospheres in a simple optical standing wave potential, for
spheres of radii $r=20-500$\,nm, illustrating a protocol for loading single
nanospheres into a standing wave optical trap that would be formed by an optical
cavity.
We use this data to confirm the dependence of the
effective optomechanical coupling strength on sphere radius for levitated
nanospheres in an optical cavity
and discuss the prospects for reaching regimes of strong light-matter coupling.
Theoretical semiclassical and quantum displacement noise spectra
show that for larger nanospheres with $r~\gtrsim~100$\,nm a range of interesting
and novel dynamical regimes can be accessed.
These include simultaneous hybridization of the two optical modes with the
mechanical modes and parameter regimes where the system is bistable.
 We show that here, in contrast to typical
 single-optical mode optomechanical systems, bistabilities are independent of
intracavity intensity and can occur for very weak laser driving
amplitudes.}
\end{abstract}
\maketitle

\section*{Introduction}

Extraordinary progress has been made in the last half-dozen years
\cite{Review,Kipp}
towards the goal of cooling a small mechanical resonator down to its
 quantum ground state and hence to realise quantum behavior
in a macroscopic system. Implementations include cavity cooling of micromirrors
on
 cantilevers \cite{Metzger,Arcizet,Gigan,Regal};
dielectric membranes in Fabry-Perot cavities \cite{membrane};
 radial and whispering gallery modes of optical microcavities \cite{Schliess}
 and nano-electromechanical systems \cite{NEMS}. Indeed the realizations
span 12 orders of magnitude in size \cite{Kipp},  up to and including the LIGO
gravity wave experiments.
In 2011 two separate experiments \cite{Teufel,Chan} achieved sideband cooling
of micromechanical and nanomechanical oscillators to the quantum ground state.
In Ref.~\cite{Asymm}, spectral signatures (in the form of asymmetric
displacement noise
spectra) of quantum ground state cooling were further investigated.
Corresponding advances in the theory of optomechanical cooling have also been
made \cite{Brag,Paternostro,Marquardt,Wilson}.

Over the last year or so, a promising new paradigm has been attracting much
interest: several groups \cite{Isart,Zoller,Barker,Ritsch,Isart2,Opto1} have now
investigated schemes
for optomechanical cooling of
levitated dielectric particles, including nanospheres, microspheres and even
viruses.
The important advantage is the elimination of the
mechanical support, a dominant source of environmental noise which can heat and
decohere the system.

In general, these proposals involve two fields, one for trapping and one for
cooling.
This may involve an optical cavity mode plus a separate trap, or two optical
cavity modes:
the so-called ``self-trapping'' scenario.

 Mechanical oscillators in the self-trapping regime differ from other
optomechanically-cooled devices in a second
 fundamental respect (in addition to the absence of mechanical support): the
mechanical frequency,
$\omega_M$, associated  with centre-of-mass oscillations is not an intrinsic
feature of the
resonator but is determined by the optical field. In particular, it is a
function of one or
both of the detuning frequencies, $\Delta_1$ and $\Delta_2$, of the optical
modes.
In previous work \cite{Opto1}, we analysed cooling in the self-trapped regime
and found that
the optimal condition for cooling occurs where both fields competitively cool
and trap the
nanosphere. This happens when $\omega_M$ is resonantly red detuned from
both the detuning frequencies i.e.
 $\omega_M(\Delta_1,\Delta_2) \sim -\Delta_{1,2}$ so the relevant resonant
frequencies
are mutually interdependent. Most significantly, the effective light-matter
coupling strength $g$  also depends on the detunings.

The effective coupling strength, ${\tilde g} = g\sqrt{n}$ (the optomechanical
coupling rescaled by the square root of photon number)
 determines whether one can attain strong coupling regimes in levitated
systems such as recently observed  in a non-levitated set-up  \cite{strong}.
It determines too whether one may access other interesting dynamics, both in the
semiclassical and quantum regimes.  We consider in particular the possibility of
simultaneous hybridization of the two optical modes with the mechanical mode;
here,
we consider also the implications  of a static bistability, which, unusually,
occurs also in the limit of quite weak
driving in the levitated self-trapped system.

In the present work, we investigate theoretically and experimentally
the strength of the optomechanical coupling.
In particular, we present experimental measurements of the mechanical frequency
of a nanosphere
trapped in an optical standing wave in order to investigate the optical coupling
as a function
of the size of the nanosphere.

In section 1 we review the theory of the cavity cooling and dynamics of a
self-trapped system,
 and in section 2 we employ the experimentally measured size dependence of the
coupling to
determine the range of optomechanical coupling strengths accessible in a cavity.
The data suggests that the most effective means to attain stronger coupling will
be
to employ larger nanospheres of typical radii $r=200-300$\,nm.
Our work suggests that  increasing
photon number by stronger driving  (and by implication increasing rescaled
coupling strengths) will not prove
an effective alternative, since in the present system we show ${\tilde g}
\propto  {n}^{1/4}$ rather
than $\sqrt{n}$, so the rescaled coupling increases very slowly with laser input
power.

In section 3, we in investigate the cooling and dynamics.
In sec.~3.1, we review the corresponding cooling rate
expressions obtained from quantum perturbation theory (or linear response
theory).
In sec.~3.2 we report a study of the corresponding semiclassical Langevin
equations
and compare them with fully quantum noise spectra; we compare also quantum,
semiclassical
and perturbation theory results  for levitated nanospheres.
 In section 4 we investigate novel
regimes  of triple mode hybridization, coincident with static bistabilities,
which the present study shows are experimentally accessible given the large
optomechanical coupling
strengths associated with $r=100-300$\,nm nanospheres.
In section 5 we describe the experimental study which provides data from which
the size-dependence of
the coupling may be inferred.
In Section 6, we conclude.

\section{Theory: Quantum Hamiltonian for a nanosphere in a cavity}

We approximate the equivalent cavity model by a one-dimensional system, with
centre-of-mass motion
confined to the axial dimension. In this simplified study, we consider only the
axial dynamics:
 for the cavity system, we will have
a much smaller tranverse frequency relative to the axial frequency, i.e.
$\omega_t \ll \omega_a$ and there is little mixing between these
transverse and axial degrees of freedom.

 We consider the dynamics of the following  Hamiltonian:
\begin{eqnarray}
\frac{\hat H}{\hbar}&=& -\Delta_1{\hat a}_1^\dagger {\hat a}_1  -\Delta_2 {\hat
a_2}^\dagger {\hat a}_2
      +  \frac{{\hat P}^2}{2m\hbar}   -A  {\hat a}_1^\dagger  {\hat a}_1 \cos^2
(k_1 {\hat x}-\phi_1)
         \nonumber \\
 &- & A{\hat a}_2^\dagger {\hat a}_2\cos^2 (k_2{\hat x}-\phi_2)
  + E_1({\hat a_1}^\dagger+ {\hat a}_1) + E_2({\hat a}_2^\dagger + {\hat a}_2).
\label{Hamiltonian}
\end{eqnarray}

Two optical field modes of a high finesse cavity ${\hat a}_{1,2}$ are coupled to
a nanosphere with
 centre-of-mass position $x$. The parameter $A$ (dependent on the nanosphere
polarizability),
determines the depth of the optical standing-wave potentials.
We investigate the case where both modes competitively cool and trap the
nanosphere, in contrast
to previous schemes \cite{Isart,Zoller} where  one optical field is exclusively
responsible for trapping,
while the other is exclusively responsible for cooling.
$\hat H$ is given in the rotating frame of the laser which drives the modes with
amplitudes
$E_1$ and $E_2=R E_1$ respectively, where $R$ represents the ratio of driving
amplitudes for
the two modes. We restrict ourselves to the regime $0 \leq R \leq 1$, since we
consider the most general case where
both optical modes contribute to the trapping as well as the cooling. Thus we
can define mode 1 simply as the mode
which is more strongly driven and mode 2 as the mode which is (except where
$R=1$) more weakly driven. The detunings
$\Delta_{j}=\omega^j_L-\omega^{j}_c$ for $j=1,2$ are between
the input lasers and the
corresponding cavity mode of interest, and $\phi_{1,2}$ represents the phases of
the optical potentials.

The two fields could represent two modes generated by the same laser field, or
they could be generated by two independent
lasers. Nonetheless, since the particle motion is confined to within one
wavelegth,
 one can make the approximation $k_1\approx k_2 \equiv k$.
Previous studies generally consider $\phi_1=0 , \phi_2=\pi/4$ to be convenient,
since then the anti-node of one field coincides with a purely linear potential
of the other optical field, but we may also consider general values of
$\phi_1-\phi_2$.
One can write corresponding equations of motion:

\begin{eqnarray}
{\ddot {\hat x}} & = & -\frac{\hbar k A}{m}\sum_j  {\hat a}_j^\dagger {\hat a}_j
\sin 2(k {\hat x}-\phi_j)
 -\Gamma_M  {\dot  {\hat x}} \nonumber\\
{\dot {\hat a}_j}& = & i \Delta_j {\hat a}_j -iE_j +i A {\hat a}_j\cos^2 (k
{\hat x}-\phi_j) -\frac{\kappa}{2} {\hat a}_j,
\label{Heisenberg}
\end{eqnarray}

where $j=1,2$ for the two optical-mode realisation.  Additional damping terms
have also been added: $\frac{\kappa}{2} {\hat a}_i$ accounts for photon losses due to mirror
imperfections and the $\Gamma_M {\dot  {\hat x}}$ term for mechanical damping.
The above should also include quantum noise terms arising from (say) shot noise or gas collisions:
for brevity, the quantum noise terms are left out until sec.3.
 
We consider here the linearised dynamics; we replace operators by their
expectation values and
perform the shifts about equilibrium values such as ${\hat a}_j(t) \to \alpha_j
+ {\hat a}_j(t)$,
and ${\hat x} \to x_0 + {\hat x}(t)$.
The values for the equilibrium photon fields (e.g. for the two-mode case) are
$\alpha_{1} = -iE_1 \left[\frac{\kappa}{2} - i\Delta_{1}^x \right]^{-1} \ $ and
$\alpha_{2}  =  -iRE_1\left[\frac{\kappa}{2}- i\Delta_{2}^x \right]^{-1}$.
The equilibrium position is then given by the relation $-\frac{\sin
2(kx_0-\phi_1)}{\sin 2(kx_0-\phi_2)} = {|\alpha_2|^2}/{|\alpha_1|^2}$,
by numerical solution of the equation,

\begin{eqnarray}
-\frac{\sin 2(kx-\phi_1)}{\sin 2(kx-\phi_2)}= R^2\frac{|\frac{\kappa}{2} -
i\Delta^x_{1}|^2}{|\frac{\kappa}{2} - i\Delta^x_{2}|^2},
\label{equil}
\end{eqnarray}

where   $\Delta_j^x=\Delta_j +A \cos^2 (kx_0-\phi_j)$.
As usual we consider the dynamics of the fluctuations via the linearised
equations.
 To first order, the linearised equations of motion, in the shifted frame, are:

\begin{eqnarray}
{\ddot {\hat x}}& = & - \omega^2_M {\hat x} - \frac{\hbar kA}{m}\sum_j
(\alpha_j^* {\hat a}_j+ \alpha_j {\hat a}_j^\dagger)\sin 2(kx_0-\phi_j)
-\Gamma_M {\dot {\hat x}} \nonumber\\
{\dot {\hat a}_j} & = & i \Delta_j^x {\hat a}_j - ikA \alpha_j {\hat x} \sin
2(kx_0-\phi_j) -\frac{\kappa}{2} {\hat a}_j.
\label{a2s}
\end{eqnarray}

The resulting effective  mechanical harmonic oscillator frequency is:

\begin{eqnarray}
\omega_M^2= \frac{2\hbar A k^2}{m} \sum_j |\alpha_j|^2 \cos{2(kx_0-\phi_j)}.
\label{freqs}
\end{eqnarray}

We can restrict ourselves to real equilibrium fields. We take $\alpha_j ={\tilde
{\alpha}}_j e^{-i\theta_j}$
then transform ${\hat a}_j \to {\hat a}_j e^{i\theta_j}$. Thus $(\alpha_j^*
{\hat a}_j+ \alpha_j {\hat a}_j^*)\equiv {\tilde {\alpha}}_j( {\hat a}_j+ {\hat
a}_j^\dagger)$.
We also rescale the mechanical oscillator coordinates ${\hat x} \to
\sqrt{2}X_{zpf}{\hat x}$ and
${\hat p} \to \sqrt{\hbar m\omega_M} {\hat p}$, where
$X_{zpf}=\sqrt{\frac{\hbar}{2 m\omega_M}} $ is
 the zero-point fluctuation length scale. Hence,
$\frac{{\hat P}^2}{2m}+\frac{1}{2} m \omega_M^2 {\hat x}^2 \to \frac{\hbar
\omega_M}{2}({\hat x}^2+{\hat p}^2)$.\\

Below we drop the tilde so the equilibrium field values ${\tilde {\alpha}}_j 
\equiv \alpha_j$ are real.
Using field operators ${\hat x}= ( {\hat b} +{\hat b}^\dagger)/\sqrt{2}$,  the
linearised dynamics
for a two-optical mode system would correspond to an effective Hamiltonian:

\begin{eqnarray}
\frac{\hat H_{Lin}}{\hbar}&=&  -\Delta_1^x{\hat a}_1^\dagger {\hat a}_1 
-\Delta_2^x{\hat a}_2^\dagger {\hat a}_2
  + \omega_M(\Delta_1^x,\Delta_2^x){\hat b}^\dagger {\hat b}\nonumber\\
     & +&  g_1(\Delta_1^x,\Delta_2^x)\alpha_1( {\hat a}_1+  {\hat a}_1^\dagger)(
{\hat b} +{\hat b}^\dagger)   +g_2(\Delta_1^x,\Delta_2^x)
\alpha_2( {\hat a}_2+  {\hat a}_2^\dagger)( {\hat b} +{\hat b}^\dagger).
\label{LinHam}
\end{eqnarray}

\begin{figure}[htb]
\begin{center}
\includegraphics[height=3.3in]{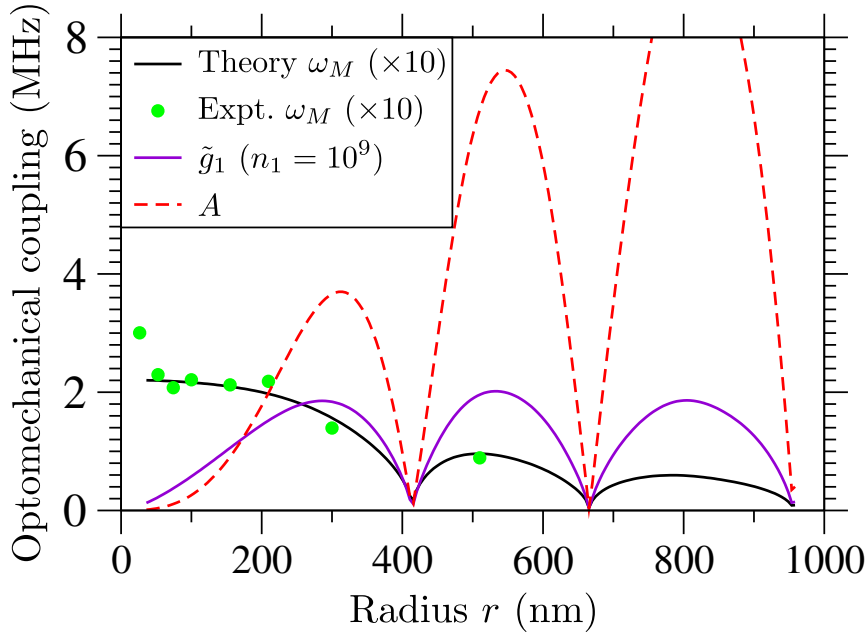}
\end{center}
\caption{(Colour online) Size dependent effects in the magnitude of the optomechanical coupling
parameter ${\tilde g}_{1}$.
It is assumed that cavity parameters would correspond to ${\tilde
g}_{1}=10^6$\,Hz at $r=150$\,nm,
for photon numbers $n_1=10^9$.
For comparison, the value of $A$ is also shown, as are the experimental and
simulated
frequencies $\omega_a \equiv \omega_M =2\pi f_M$.
The $\omega_M$ are scaled by a factor of 10 for clarity and in fact, values of
$\omega_M \sim 1$\,MHz are
quite realistic in optical cavities.}
\label{beadg}
\end{figure}

\section{Towards strong light-matter coupling with two optical cavity  modes}
The Hamiltonian in Eq.~\ref{LinHam} appears analogous in form to standard,
well-studied optomechanical Hamiltonians, albeit with two optical modes rather
than one. However, it
differs in one important respect: in this case,
 both the  mechanical frequency $\omega_M(\Delta_1^x, \Delta_2^x)$ and the
optomechanical
 coupling strengths $g_{1,2} \equiv g_{1,2}(\Delta_1^x,\Delta_2^x)$ are not
fixed and depend on the detunings.
 The fact that the frequencies of the three modes (two optical, one mechanical)
are interdependent makes the dynamics different from other optomechanical
set-ups, where the equilibrium mechanical
frequency (i.e. excluding shifts arising from the fluctuations)
 is intrinsic to the mechanical oscillator.

There is considerable interest in achieving strong-coupling, which leads to
regimes of light-matter
hybridization. The corresponding mode splitting has been observed experimentally
\cite{strong}.
In typical set-ups, these regimes are reached if the rescaled effective
optomechanical
coupling exceeds the damping rates i.e. ${\tilde g} = g \sqrt{n} \gtrsim \kappa,
\Gamma_M$, where
$n \sim |\alpha|^2$ is the cavity photon number. Even if the unnormalized
coupling is weak, strong-coupling
regimes may be achieved by increasing the driving power and thus increasing
intracavity photon numbers.

In the present levitated system, a particularly interesting regime would involve
triple-mode hybridization
  enabling, for example, the coupling of
the two modes of light via the mechanical mode. However, here, mode
hybridization
(for which ${\tilde g}_{1,2}= g_{1,2} \alpha_{1,2} \gtrsim \kappa, \Gamma_M$)
depends
non-trivially on the detunings.

We argue that large coupling cannot be easily achieved by increasing the
driving power, since  ${\tilde g}_{1,2} \propto n_{1,2}^{1/4}$ and thus
increases slowly with the
driving strength. We can show, by a simple argument, that increasing the
nanosphere size provides the most
effective means to attain strong coupling.

For the self-trapped system, optomechanical coupling strengths are $g_j=
\sqrt{2}kA X_{zpf} \sin 2(kx_0-\phi_j)$ and
 depend  on the detunings via $x_0$. Note also that $
X_{zpf}=\sqrt{\hbar/(2m\omega_M)}$ here
too depends on the detunings via $\omega_M$.

For triple mode hybridization,  $\omega_M \sim \Delta_{1} \sim \Delta_{2} $.
For convenience, we also take $\phi_1=0, \phi_2=\pi/4$. Then,
since $\tan 2kx_0 = {|\alpha_2|^2}/{|\alpha_1|^2}$, we can re-write
Eq.~\ref{freqs}:

\begin{eqnarray}
\omega_M^2= \frac{2\hbar A k^2}{m \cos{2kx_0}}|\alpha_1|^2.
\label{freq2}
\end{eqnarray}

We consider near symmetric driving
of the two optical modes for which $R \sim 1$ and
 thus $kx_0 \approx \phi/2=\pi/8$ so  $\omega_M \sim  \left({\frac{2\hbar A
k^2}{m}}\right)^{1/2} n_{1,2}^{1/2} $.
Hence,

\begin{eqnarray}
{\tilde g}_{1,2} \sim \left( {\frac{\hbar k^2}{4}}\right)^{1/4} \left( \frac{A^3
n_{1,2}}{ m}\right)^{1/4}.
\end{eqnarray}

Since the optomechanical coupling increases only very slowly with cavity photon
number the most effective means to reach strong coupling regimes is to increase
the nanosphere size
to the maximum practical value ($r\sim 200-250$\,nm).

For the ideal case where the nanosphere radius $r$ is small \cite{Zoller} ie.
$\lambda \gg r$, then
$A(r\ll \lambda) \equiv A_0(r)$ where the small nanosphere coupling takes the
form:
\begin{eqnarray}
A_0(r) = \frac{3}{2} \frac{\epsilon_r-1}{\epsilon_r-2} \frac{V_s}{V_c} \omega_L
\label{ideal}
\end{eqnarray}
where $V_s= 4/3\pi r^3$ is the sphere volume (and hence $m=V_s \rho$ where the
density
$\rho=2000$\,Kg\,m$^{-3}$ for silica). In turn, $V_c=\pi (w/2)^2 L$ is the
cavity volume, where $w\approx 40\,\mu$m is the cavity
waist and $L \simeq 0.5 - 1$\,cm is the cavity length.

For larger nanospheres, the measured size-dependent corrections must be applied.
In the experiments described below, we find that the mechanical oscillation
frequency is modulated by a
finite size correction $\omega_M(r) = \omega_M(r\simeq 0) f(r)$ (see 
Fig.~\ref{exp_trap_freq} and
description of the measurement of $f(r)$ in section~5 below).
Thus, since:
\begin{eqnarray}
\omega_M^2 \propto \frac{A(r)}{m},
\label{freq2}
\end{eqnarray}

then $A(r)\equiv A_0(r) f^2(r)$ and the coupling is in turn modulated by
the finite size correction. The experimental results suggest that for
$r\lesssim 200$\,nm, then $f(r)\sim 1$ and $A(r) \simeq A_0(r)$.

 For example, for  $r=150$\,nm, $L=1$\,cm and $w=40\,\mu$m, then $A_0\simeq 8
\times 10^5$\,Hz.
For reasonable values of cavity decay constants $\kappa \simeq 2-8 \times
10^5$\,Hz, then
for $n_1 \sim 10^{9}$,
\begin{eqnarray}
{\tilde g}_{1,2} \sim 5.4 \times 10^{-6} \left({\frac{A^3
n_{1,2}}{m}}\right)^{1/4} \simeq 10^6\,\mathrm{Hz} \gg \kappa
\end{eqnarray}

For $r \lesssim 200$\,nm, ${\tilde g}_{1,2} \propto r^{3/2}$. Thus a 200\,nm
sphere
provides an optomechanical coupling about an order of magnitude larger than a
40-50\,nm sphere.
To achieve a comparable increase in coupling by photon number enhancement
would require increasing the driving power by a factor of order $10^4$.

A more careful analysis, including the effects of the finite-size correction
function $f^2(r)$
is shown in Fig.~\ref{beadg}. We see that ${\tilde g}_{1}$ (and for $n_1\sim
n_2$, also ${\tilde g}_{2}$)
 reaches a maximum value for $r\simeq 300$\,nm before falling to zero. Other
maxima for larger $r$
do not provide a larger value of ${\tilde g_{1,2}}$. Furthermore, they have the
disadvantage that they may enhance
photon recoil heating effects.
For comparison, the value of $A$ is also shown, overlaid on the experimental and
simulated
frequencies.

\section{Dynamics}
\subsection{Optomechanical damping}

 A previous study \cite{Opto1}, using rescaled coordinates, investigated the
full
parameter space of two optical mode cooling. Here we investigate more carefully
the effect of non-zero mechanical damping.
Using linear response theory, we can extract cooling rates from the equations
Eq.~\ref{a2s}:

\begin{equation}
\Gamma_{opt}=
\left[S_1(\omega_M)+S_2(\omega_M)-S_1(-\omega_M)-S_2(-\omega_M)\right],
\label{gamma}
\end{equation}

where

\begin{equation}
S_j(\omega)=\frac{ |\alpha_j|^2 g^2_j \kappa}{[\Delta_j^x-\omega]^2 +
\frac{\kappa^2}{4}},
\label{gamma12}
\end{equation}

for $j=1,2$. Net cooling occurs for $\Gamma_{opt} <0$. Although the above is
quite similar in form to standard optomechanical
expressions, as explained previously, rather different behaviour is observed
since here
$\omega_M$ and $g_j$ are both dependent on the $\Delta_j^x$.

From quantum perturbation theory we can show that $R_{n\to m}$, the rate of
transition from state $n$ to $n+1$ is:
$R_{n \to n+1}=(n+1) \left( S_1(\omega_M)+S_2(\omega_M) \right)$
while
 $R_{n \to n-1}=n  \left( S_1(-\omega_M)+S_2(-\omega_M) \right)$

For $n>>1$, then $R_{n \to n+1}- R_{n \to n-1}$ gives the cooling rate of
Eq.\ref{gamma}.  However, with the
 exact expressions we can show that the equilibrium mean phonon number is

\begin{equation}
\langle n
\rangle_{PT}=\frac{S_1(\omega_M)+S_2(\omega_M)}{
S_1(-\omega_M)+S_2(-\omega_M)-S_1(\omega_M)-S_2(\omega_M)}.
\label{neq}
\end{equation}

\begin{figure}[htb]
\begin{center}
\includegraphics[height=3.3in]{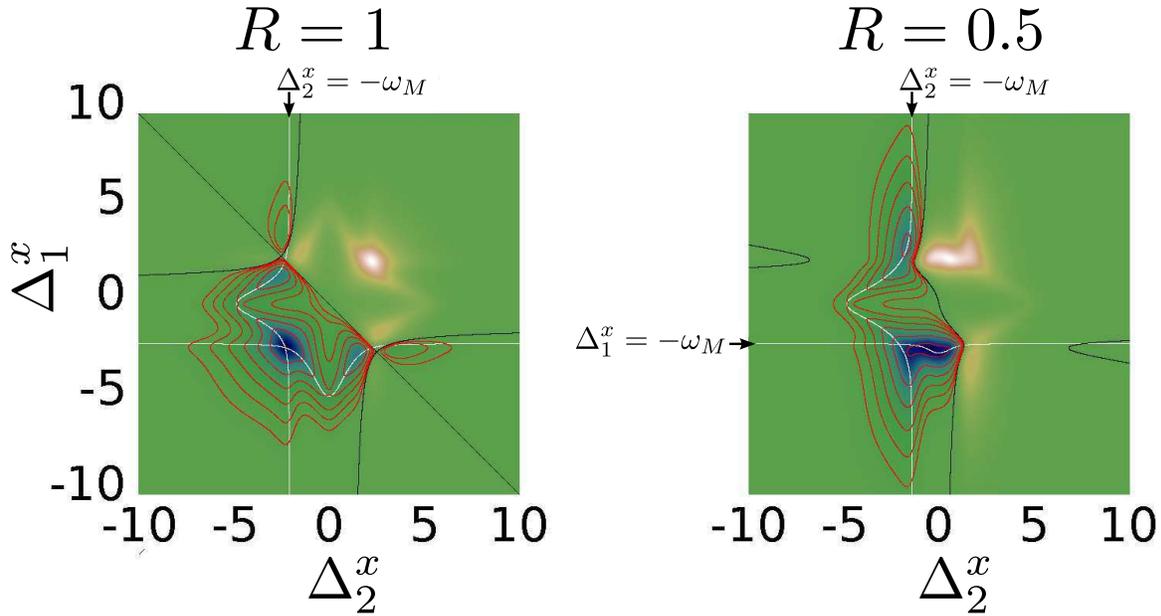}
\end{center}
\caption{(Colour online) Maps of cooling rates calculated from Eq.~\ref{gamma}
for parameters $R=1.0$ and $R=0.5$. Blue corresponds to cooling, yellow/white to heating.
The white lines indicate the locus of the single field resonances (where
$-\Delta_{1}^x=\omega_M$
 or where $-\Delta_{2}^x=\omega_M$).
The detunings are given in units of $A$ and are dimensionless.
 For $R=1$ it is clear that there is a deep, maximum
cooling region at a double resonance where the two white lines intersect and
both optical fields
cool simultaneously. It is also evident that there is a strong cooling resonance
for $+\Delta_{1,2\pm}^x=\omega_M$.
For $R=0.5$, three cooling resonances $-\Delta_{1\pm}^x=\omega_M$,
$-\Delta_{2\pm}^x=\omega_M$
and  $+\Delta_{1}^x=\omega_M$ merge to give a  very broad strong-cooling {\em
region}, quite insensitive to detuning $\Delta_2$  over a range of over
1\,MHz.
Here $A=\kappa/2= 0.3$\,MHz and the input power into mode 1 corresponds to
2\,mW.}
\label{Cool}
\end{figure}

In Fig.~\ref{Cool}  we show colour maps comparing the cooling
and minimum phonon numbers for both $R=0.5$ and $R=1$ respectively. 
The cooling behaviour was investigated previously in \cite{Opto1}.
In this case, for each fixed detuning $\Delta_1$ there are up to three
cooling resonances (at three different values of $\Delta_2$),
 where strong damping is observed (and similarly for each fixed $\Delta_2$).
This is in contrast to  single optical mode schemes where there is a single
cooling resonance for which $\Delta_{1}=-\omega_M$ or $\Delta_{2}=-\omega_M$.
For the $R=0.5$ map the three cooling resonances merge, giving a single extended
cooling region of about 1\,MHz width.

For $R=1$ the map has a
high degree of symmetry, since  the role of the two optical modes is
interchangeable.
The figures show that the largest cooling rates are found in the double
resonance region, making it the most favourable region to work in.

The equilibrium phonon number in Eq.~\ref{neq} concerns only the idealised
situation where
there is a very good vacuum, negligible photon recoil heating and thus no mechanical
damping or heating effects.
For small $r \lesssim 200$\,nm spheres, we assume recoil heating is negligible
\cite{Zoller} and the dominant
source of mechanical damping is background gas collisions, which provide an
effective
mechanical damping $\Gamma_M=\frac{8}{3} \pi \frac{m_g}{m_s}  r^2 n_g \bar{v}_g$
\cite{Isart,Zoller}
 where $m_g / m_s$ is ratio of the gas particle's mass to that of the sphere,
$n_g$ is the gas number density and $\bar{v}_g$ is the mean gas velocity
for a room temperature thermal distribution.

It can be shown that the perturbation theory argument above can be adapted to
obtain equilibrium phonon numbers
for a given cooling rate $\Gamma_{opt}$:

\begin{eqnarray}
\langle{n}\rangle_{PT}=\frac{\frac{k_BT_B}{\hbar\omega_M} {\Gamma_M} + 
\left[S_1(\omega_M )+
S_2(\omega_M )\right]}
{\Gamma_M + |\Gamma_{opt}|},
\label{PT}
\end{eqnarray}

where $T_B\simeq 300$K.
Alternatively, the final equilibrium temperatures $T_{eq} = \frac{\Gamma_M T_{B}
+ |\Gamma| T_{vac}}{\Gamma_M+|\Gamma|}$,
where $T_{vac}$ is the equilibrium oscillator temperature which would have been
obtained in a perfect vacuum.

\subsection{Quantum and semiclassical noise spectra}
Although we investigate only a two optical mode system, generalization to more
optical modes
is straightforward.
We consider a set of equations of motion, for $j=1,...N$:
\begin{eqnarray}
 { \dot { \hat b} } & = &  -( i\omega_M (\Delta_1^x,...\Delta_j^x)+
\frac{\Gamma_M}{2} ) {\hat b} +
i\sum_j g_j(\Delta_1^x,...\Delta_j^x) ({\hat a}_j+{\hat a}_j^\dagger)   +
\sqrt{\Gamma_M} {\hat b}_{in}    \nonumber \\
{\dot {\hat a}_j} & = & (i \Delta_j^x-\frac{\kappa}{2}) {\hat a}_j + i 
g_j(\Delta_1^x,...\Delta_j^x)\alpha_j({\hat b}+{\hat b}^\dagger)
   + \sqrt{\kappa}{\hat a}_{in}^{(j)}, \nonumber\\
\label{QM}
\end{eqnarray}

where the optomechanical strengths $g_j(\Delta_1^x,...\Delta_j^x)= -kA X_{ZPF}
\sin 2(kx_0-\phi_j)$ depend on
the detunings (as does the mechanical frequency $\omega_M$).
In the two mode case we consider, we take $\phi_1=0$ and $\phi_2=\pi/4$.

The optical modes are subject to photon shot noise, while the mechanical modes
are subject to Brownian noise
from collisions with gas molecules in the cavity. For the photon shot noise, we
assume independent lasers and
 uncorrelated zero temperature noise
for which $ \langle {\hat a}_{in}^\dagger(t') {\hat a}_{in}(t)\rangle =0$, while
$ \langle {\hat a}^{(i)}_{in}(t') {\hat a}^{(j)\dagger }_{in}(t)\rangle
=\delta(t-t')\delta_{ij}$.
 For the gas collisions, we take $ \langle {\hat b}_{in}(t') {\hat
b}_{in}^\dagger(t)\rangle =(n_B+1)\delta(t-t')$
and $ \langle {\hat b}_{in}^\dagger(t') {\hat b}_{in}(t)\rangle=
n_B\delta(t-t')$ where the number of surrounding bath
phonons $n_B \approx \frac{k_BT}{\hbar \omega_M}$.

The above equations can be integrated in frequency space to obtain analytical
expressions for the displacement noise spectra
for the arbitrary mode case. We can evaluate the displacement spectrum 
$S_{xx}(\omega) \equiv \langle |x(\omega)|^2\rangle_{QM}= \frac{1}{2\pi} \int e^{-i\omega \tau} \langle
x(t+\tau) x(t)\rangle d\tau$. 
We obtain:

\begin{eqnarray}
\langle |x(\omega)|^2\rangle_{QM} |M(\omega)|^2 & = & {\Gamma_M} \left[
|\chi_M(\omega)|^2 n_B + |\chi_M(-\omega)|^2 (n_B+1)\right]\nonumber \\
                                               & + & \frac{\kappa}{2}
|\mu(\omega)|^2 \sum_{j=1,2} g_j^2 |\chi_{jo}(-\omega)|^2,
\label{qnoise}
\end{eqnarray}

where the $\chi(\omega)$ represent optical and mechanical susceptibilities:
\begin{eqnarray}
\chi_{jo}(\omega)=\left[-i(\omega+\Delta_j^x)+\frac{\kappa}{2}\right]^{-1};  \
 \chi_M(\omega) =\left[-i(\omega-\omega_M)+\frac{\Gamma_M}{2}\right]^{-1}.
\end{eqnarray}
 with $ \mu(\omega)=\chi_M(\omega)-\chi_M^*(-\omega)$ and $ \eta_j(\omega)=
\chi_{jo}(\omega)-\chi_{jo}^*(-\omega)$;
then also $M(\omega)= 1+ \mu(\omega)\sum_j g_j^2 |\eta_{j}(\omega)|^2 $.

\begin{figure}[h]
\begin{center}
\includegraphics[height=3in]{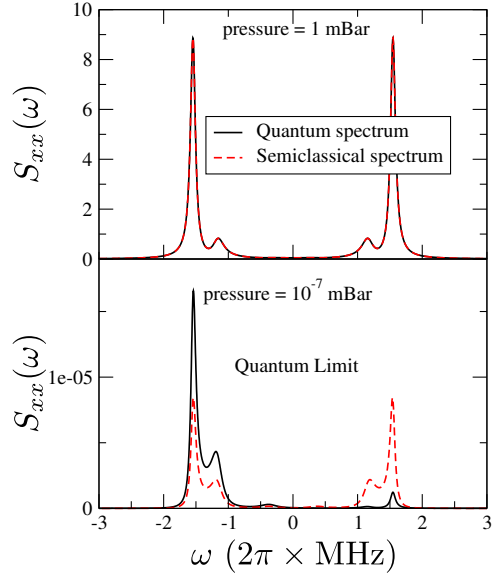}
\end{center}
\caption{(Colour online) Comparison between quantum and semiclassical
displacement spectra for gas pressures
of 1\,mBar and at near vacuum pressure in a strong cooling region. At high
vacuum, the ground state is
approached and thus, for the quantum spectrum, the blue sideband vanishes.
At higher pressure there is good agreement between the quantum and classical
results.
Spectra near double resonance for input power $P_1=7$\,mW, $A=\kappa=3\times
10^5$\,Hz,
$\Delta_1=-1.5$\,MHz, $\Delta_2=-0.68$\,MHz, $R=0.5$. Some hybridization between
the mechanical
mode and optical mode 1 is seen in the characteristic double-peak sideband
structure.}
\label{Noise}
\end{figure}

\begin{figure}[h]
\begin{center}
\includegraphics[height=3in]{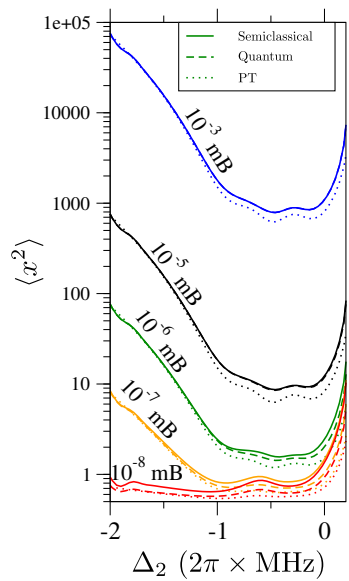}
\end{center}
\caption{For the broad cooling region formed from three overlapping resonances
seen in Fig.~\ref{Cool}(b),
we show equilibrium phonon numbers obtained from  Eqs.~(\ref{PT}),
(\ref{qnoise}) and (\ref{SSC})
i.e. perturbation theory,
the analytical quantum noise formula and  semiclassical Langevin equations
respectively.
Agreement between quantum and semiclassical results is excellent, less so for 
perturbation theory at low pressures.}
\label{Phonon}
\end{figure}

\begin{figure}[h]
\begin{center}
\includegraphics[height=3.5in]{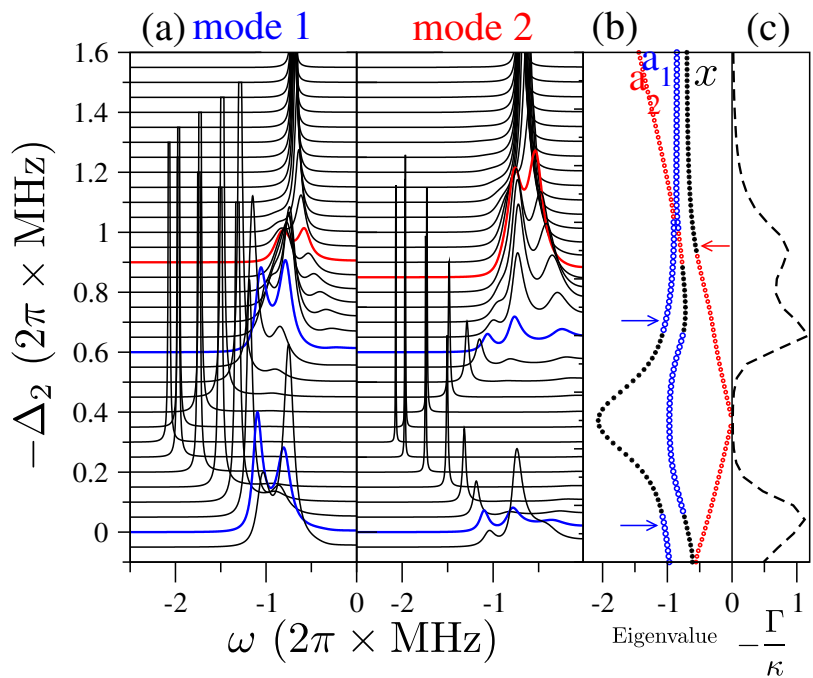}
\end{center}
\caption{Triple mode splitting. For $A=\kappa=3$\,MHz, even at quite high
pressures (here 1\,mBar), mode
splitting is seen in the noise spectra of the optical modes.
In all the plots,  $\Delta_1 = -1.15$\,MHz is held fixed while
 $\Delta_2$ is swept from 0 to -1.6\,MHz
(for an input power of  2\,mW into mode 1, while $R=0.5$).
(a) Shows noise spectra  for both optical mode 1 and mode 2.
Three way hybridization between the mechanical and both optical modes
appears clearly (highlighted in the bold blue line). For clarity,
some of the strongest peaks have been truncated in height.
 In (b) three avoided crossings are apparent.
The dominant character of each  normal mode is indicated by the colour
(black is mechanical,  blue is optical mode 1, red is optical mode 2).
When $-\Delta_2$ is large, there is no mixing. However as $\Delta_2 \to 0$,
there is strong mixing and the dominant character of each normal mode changes
from light to matter (or vice versa) as an avoided crossing is encountered.
 Panel (c) shows the cooling and indicates
strong cooling at each of the avoided crossings. }
\label{eigensplit}
\end{figure}

We compare the quantum displacement with corresponding semiclassical solutions
in the steady state.
The linearised two mode system Eq.~\ref{LinHam}, in matrix form corresponds to a
standard problem \cite{Milburn}.
 Inclusion of the noise arising from gas collisions or laser shot
noise yields a set of corresponding Langevin equations:
$ \frac{d{\bf X(t)}}{dt}= {\bf A}{\bf X}+ {\bf B E}(t)$,
where ${\bf A}$ is termed the drift matrix. Its eigenvalues give the stabilities
and eigenfrequencies of the
system's normal modes, while the noise is determined by
${\bf B}$, a  constant diagonal  matrix.
The elements of the random noise matrix are assumed to be  $\delta$-correlated 
$\langle E_i(t) E_j(t')\rangle= \delta(t-t') \delta_{ij}$.
Methods for obtaining the solution for the steady state correlation functions of
this system, under conditions of stability, i.e. where all the eigenvalues of
${\bf A}$ have negative real parts, are well-known \cite{Milburn}.
The required noise spectra, or autocorrelation functions, in frequency space
are:

\begin{eqnarray}
{\bf S}(\omega)=\frac{1}{2\pi} _{-\infty}\int^{\infty} e^{-i\omega \tau} \langle
{\bf X}(t+\tau) {\bf X^T}(t)\rangle d\tau,
\end{eqnarray}

where,

\begin{eqnarray}
{\bf S}(\omega)=\frac{1}{2\pi} \left({\bf A}+ i \omega{\bf I}\right)^{-1}
               {\bf B}{\bf B^T}\left({\bf A^T}- i \omega{\bf I}\right)^{-1},
\label{SSC}
\end{eqnarray}

where the diagonal matrix ${\bf B}{\bf B^T}$ has elements
 $\left((n_B+\frac{1}{2}){\Gamma_M},
(n_B+\frac{1}{2}){\Gamma_M},\frac{\kappa}{2},\frac{\kappa}{2},\frac{\kappa}{2},
\frac{\kappa}{2}\right)$.
From the above, the noise spectra of all modes may be calculated.  Eq.{SSC}
yields semiclassical sideband spectra, symmetrical in $\omega$.

In Fig.~\ref{Noise}, we compare semiclassical displacement spectra calculated
from Eq.~\ref{SSC} with corresponding
quantum results obtained from Eq.~\ref{qnoise}. At high pressures (and hence
high phonon occupancy)
there is excellent agreement between semiclassical and quantum results. At low
pressures (near ground
state cooling) however, the quantum spectrum shows a characteristic asymmetry,
such as was observed recently
in experiments on photonic cavities \cite{Asymm}.

\begin{figure}[h]
\begin{center}
\includegraphics[height=3in]{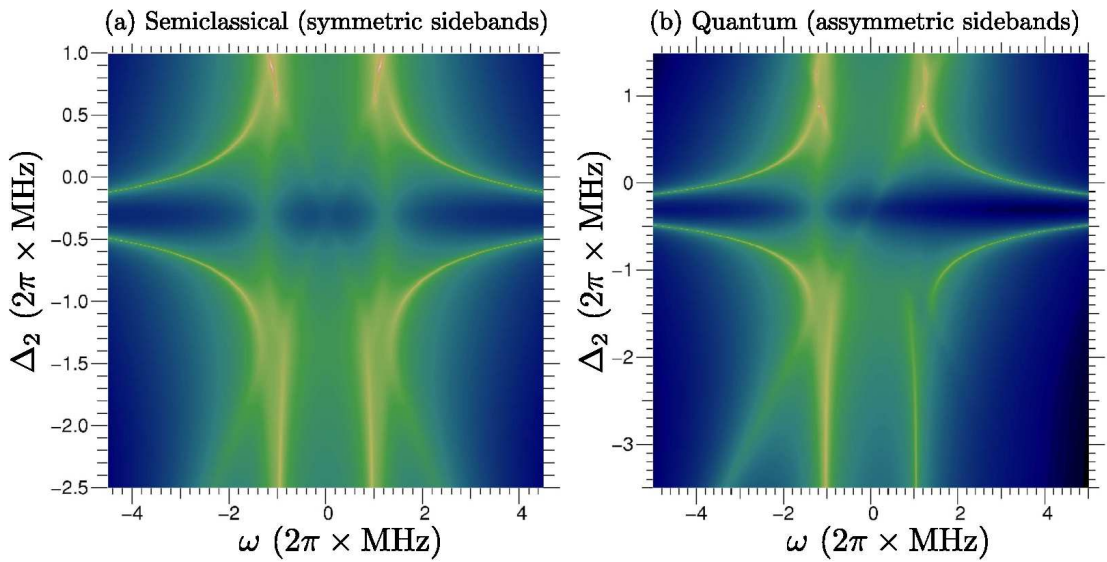}
\end{center}
\caption{Maps of the displacement noise spectra $S_{xx}(\omega)$, showing mode
splitting for similar parameters to Fig.~\ref{Noise},
near the quantum limit, except that here  $\Delta_1 = -1.5$\,MHz is held fixed
while $\Delta_2$ (vertical axes) is swept. (a) Shows the semiclassical
spectrum which is symmetric in frequency 
$\omega$. (b) Shows the quantum spectrum which is asymmetric. In both
cases, triple hybridization appears clearly
and is seen near  $\Delta_2 \approx -1.0$\,MHz. (c) shows the
corresponding behaviour for $A=\kappa/2$, showing that
the triple peak structure has disappeared. Note that $\log{ S_{xx}(\omega)}$ is
plotted.}
\label{SPLIT}
\end{figure}

\subsection{Comparison between perturbation theory, semiclassical and quantum
results }

The equilibrium variance (and hence the final phonon number) of the mechanical
oscillator is,

\begin{eqnarray}
\langle x^2\rangle=\frac{1}{2\pi} _{-\infty}\int^{\infty} \langle
|x(\omega)|^2\rangle d\omega,
\end{eqnarray}

thus, the final equilibrium temperature of the mechanical oscillator after
optomechanical cooling is $k_BT_{eq}=1/2m \omega_M^2 \langle X^2\rangle$.
Noting the rescaling $ \langle X^2\rangle= 2X_{ZPF} \langle x^2\rangle$ and
setting  $k_BT_{eq}=(\langle n \rangle +1/2) \hbar \omega_M$,
we can write  $\langle x^2\rangle= \langle n \rangle+1/2$.

Using Eqs.~(\ref{QM}), (\ref{SSC}) and (\ref{PT}), we can investigate final
equilibrium phonon numbers
(and the minimum achievable for levitated self-trapped spheres) comparing
quantum, semiclassical and perturbation
theory respectively.
In Fig.~\ref{Phonon}, we compare the corresponding equilibrium phonon numbers,
$\langle n \rangle_{QM}$, $\langle n \rangle_{SC}$
and $\langle n \rangle_{PT}$ respectively for the unusual triple cooling
resonance region shown in Fig.~\ref{Cool}.
Cooling to near the ground state $ \langle n \rangle \sim 0$ is possible for a
pressure of order $10^{-6}$\,mBar, even
for modest driving powers of 2\,mW and values of $A \simeq 3 \times 10^5$\,Hz
corresponding to spheres of order $r \simeq 100$\,nm.

\begin{figure}[ht]
\begin{center}
\includegraphics{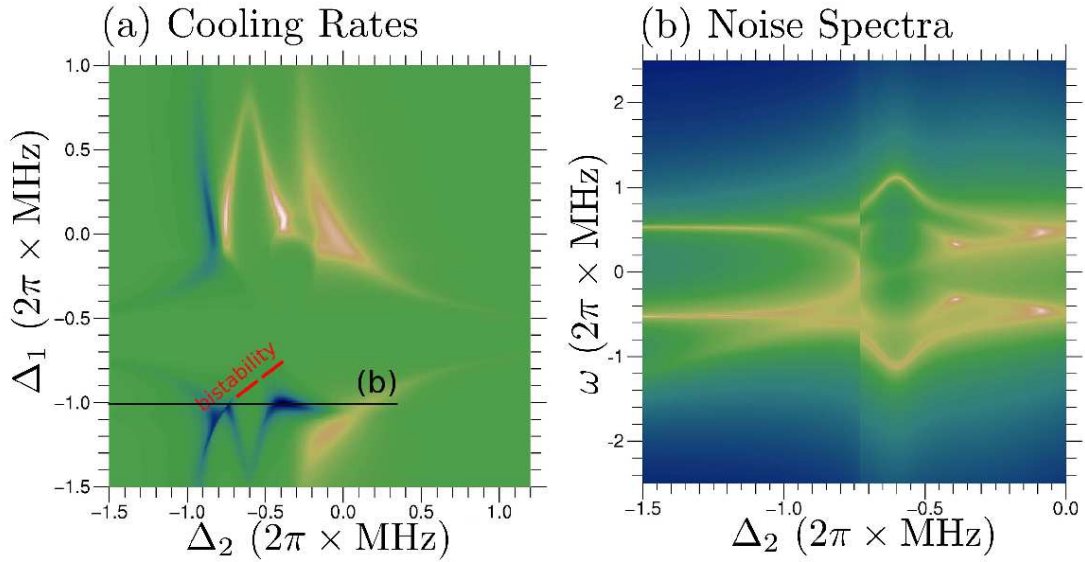}
\end{center}
\caption{Mode mixing and bistability for $A=3\kappa=6 \times 10^{5}$\,Hz.
We consider a relatively low input power of 0.37\,mW. (a) plots the
optomechanical cooling rate (blue indicates
cooling, brown indicates heating). The red dashed line indicates the locus of
bistability as a function of the detunings
$\Delta_1$ and  $\Delta_2$. The discontinuity in the cooling can be discerned
near the strong doubly resonant
cooling region. (b) shows displacement noise spectra $S_{xx}(\omega)$ as a
function of $\omega$
plotted along $\Delta_1=-1$\,MHz (along black horizontal line in upper panel).
We sweep in increasing $\Delta_2$.
At the point where  $\Delta_1=-1$\,MHz intersects the bistability, a
discontinuity
in the noise spectra is apparent.
On one side of the discontinuity there is strong hybridization between the
mechanical mode
and optical mode 2; this changes abruptly across the discontinuity to
hybridization between the
mechanical mode and optical mode 1 for larger  $\Delta_2$. This may allow for
control of
entanglement between the modes. The map corresponds to near-vacuum conditions so
the system is near
the quantum ground state in this regime (as evidenced by the asymmetric
sidebands).}
\label{bi}
\end{figure}

\section{Strong coupling regimes: triple mode splitting and bistability}

The multimode, or at least two mode, self-trapping regime may permit  new
possibilities for
position sensing and for controlling entanglement between two optical modes and
the mechanical resonator.
Here we investigate regimes where such effects are clearly apparent. The
implications
of the measurement for the accessible range of optomechanical coupling strengths
suggests that  multiple hybridization and bistability are quite accessible with
reasonable cavity  parameters.

In Fig.~\ref{eigensplit} we investigate the complex behaviour of the eigenmode
frequencies
of the self-trapped, levitated system. On the left panels (Fig.~\ref{eigensplit}
(a)) we plot the
noise spectra of the two optical modes, which exibit sidebands near $\omega
\simeq \omega_M$
since the corresponding optical fields are
modulated by the motion of the mechanical oscillator. Here we fix one detuning
($\Delta_1=-1.15$\,MHz) and
look at the behaviour as the other detuning is varied.
The sidebands are displaced in frequency and split: one effect is simply due to
the dependence of  $\omega_M$ on $\Delta_j$ (unique to the levitated system);
it arises from the calculation of the equilibrium fields and frequencies.
 The other effect is due to normal mode mixing (hybridization of light and
matter modes)
arising from the linearised equations.
If $\omega_M \simeq \Delta_1 \simeq \Delta_2$ simultaneous hybridization is
observed, provided
${\tilde g}_{1,2} \gtrsim \kappa$. Figure~\ref{eigensplit} (b) shows that there
are several distinct
avoided anti-crossings, where the dominant character of each
eigenmode changes; if two crossings coincide, the spectra show a characteristic
triple-peak structure
(symmetric about $\omega=0$ in the semiclassical regime shown here). Panel (c)
shows that the corresponding
cooling rate is enhanced at each avoided level crossing.

In Fig.~\ref{SPLIT} we plot displacement spectra corresponding to
Fig.~\ref{Noise}, but over a range of
values of $\Delta_2$. A log-scale is used for $S_{xx}(\omega)$.
 The triple mixing which can
appear when two avoided crossings nearly coincide is clearly apparent at
$\Delta_2 \simeq 1$\,MHz.

Static bistability in a cavity of varying length has been seen experimentally
\cite{bistab}.
The potential for generating entanglement has recently been investigated in an
optomechanical system \cite{Ghobadi};
 however,  a relatively high laser power $P \sim 50$\,mW is required.

For the self-trapped systems, the incoherent sum of the optical standing-wave
potentials $\cos^2 (kx-\phi_1)$ and $\cos^2 (kx-\phi_2)$
does not by itself produce a double-well structure; nevertheless, as we see
below, in combination
with optomechanical shifts, bistabilities are observed, even for weak driving.
 Whether a double-well structure emerges, or not, is completely independent of
the driving power (where $ P \propto E_1^2$)
and can emerge at very low input powers, as we demonstrate below.
It is easy to see that the levitated  particle moves in an effective static
potential $V(x)$ where:

\begin{eqnarray}
\frac{dV(x)}{dx}= \hbar k AE_1^2 \left[\frac{\sin 2(kx-\phi_1)}{|(\kappa/2) -
i\Delta_{1}(x)|^2}+
                               R^2\frac{\sin 2(kx-\phi_2)}{|(\kappa/2) -
i\Delta_{2}(x)|^2}\right], \nonumber\\
\label{pot}
\end{eqnarray}

and

\begin{eqnarray}
V(x)= \hbar A E_1^2\left[\tan^{-1}
\left({\frac{\Delta_{1}(x)}{\frac{\kappa}{2}}}\right)+
                 R^2\tan^{-1}\left({\frac{
\Delta_{2}(x)}{\frac{\kappa}{2}}}\right)\right]. \nonumber \\
\label{pot1}
\end{eqnarray}

Here, note that the shifted detuning $\Delta_{j}(x)=\Delta_j+A \cos^2
(kx-\phi_j)$ is dependent on
$x$, not the equilibrium displacement $x_0$.
This potential admits two stable equilibrium points over parameter regimes where
$A \gg \kappa$
(in practice, such bistability is observed already for $A/\kappa \simeq 3$).
It is evident that the driving power factors out, so does not affect the shape
of the potential,
providing only a scaling factor.
We show in Fig.~\ref{bi} that for a high $A/\kappa$ ratio simultaneous
hybridisation and bistability co-exist:
we show that that for $P=0.37$\,mW , $A=3\kappa$ , $R=0.15$  and modest photon
numbers $n_1 \sim 10^8$ we can switch discontinuously from
hybridisation between the mechanical mode and optical mode 1, to hybrization
between the mechanical mode and  mode 2.
We take $\phi_1=0, \phi_2=\pi/4$. In the noise spectra, the switch is heralded
by a large zero-frequency peak in the displacement spectra, which is
clearly apparent in Fig.~\ref{bi}.

\section{Experiment}
\subsection{Current experimental status: Loading protocols and variation of trap
frequency with radius}

\begin{figure}[ht]
\begin{center}
\includegraphics{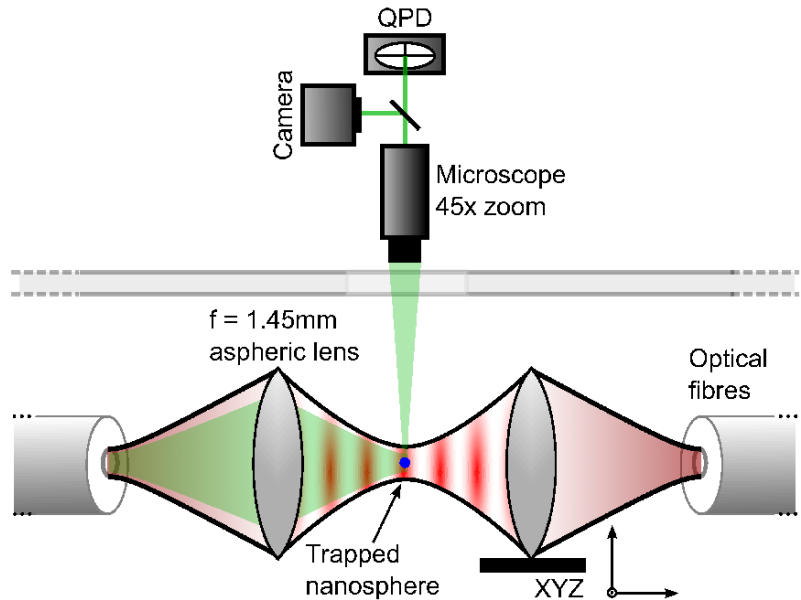}
\end{center}
\caption{(Colour online) Schematic diagram of the standing wave trap. It is
formed from two counter-propagating 1064\,nm beams focused inside a vacuum
chamber. Light at 532\,nm enters via one fibre to image the sphere. Images and
measurement of the axial and transverse position of the trapped nanosphere as a
function of time are measured by a CCD camera and quadrant cell photodiode (QPD)
respectively, through a long working distance microscope outside the vacuum
chamber.}
\label{trap}
\end{figure}

We have built a simple standing wave dipole trap to develop protocols for
loading a single nanosphere into the trap to confirm that nanospheres with a
range of radii around 100\,nm can be trapped. Importantly, we have measured the
variation in trap frequency with sphere radius so that realistic values of
optomechanical coupling strengths,
 for a given nanosphere radius, can be included in our models. In this section
we explain how the
size-dependent modulation function $f(r)$, used in the theory section above to
obtain $A(r)$ and the optomechanical
coupling strengths, was measured.

  A schematic of the standing wave trap used in our experiments  is shown in
Fig.~\ref{trap}.
 The standing wave trap consists of two focused beams that counter-propagate and
overlap near their foci. The two laser beams, derived
from the same laser at a wavelength of 1064\,nm, enter the trapping region via
optical fibres. The light exiting the fibres is focused
 using aspheric lenses (Thorlabs C140TME) with a focal length of 1.45\,mm and
numerical aperture 0.55. The power in each trapping beam
after it has passed through each lens is measured to be $150\pm10$\,mW, and the
best focused beam waist  (radius) is theoretically 1.7\,$\mu$m.
 To optimize the alignment of the trap, we maximize the light through-coupled
from one fibre into the other.  This is accomplished by mounting one
 optical fibre and its aspheric lens on an XYZ flexure stage. The alignment is
done inside the vacuum chamber at atmospheric pressure.

A long working distance microscope (Navitar Zoom 6000 system, with up to 45x
zoom) is used to image the trapped sphere. The image is split into
 two using a beamsplitting plate, with one image directed to a CCD camera for
diagnostics and the other aligned onto a quadrant cell
photodiode (QPD) which measures position fluctuations as a function of time in
two orthogonal axes. We define the axial direction as
 that along which the trapping light propagates, and the transverse direction as
the orthogonal axis in the focal plane of our
imaging system. Light at 532\,nm is used to illuminate the trapped sphere, as
the QPD is more sensitive at this wavelength.
 The green beam enters the system via one of the optical fibres, as shown in
Fig.~\ref{trap}, and a filter is used to
 stop 1064\,nm light reaching the detectors. The power of the 532\,nm beam is
10~mW.

\begin{figure}[htb]
\begin{center}
\includegraphics{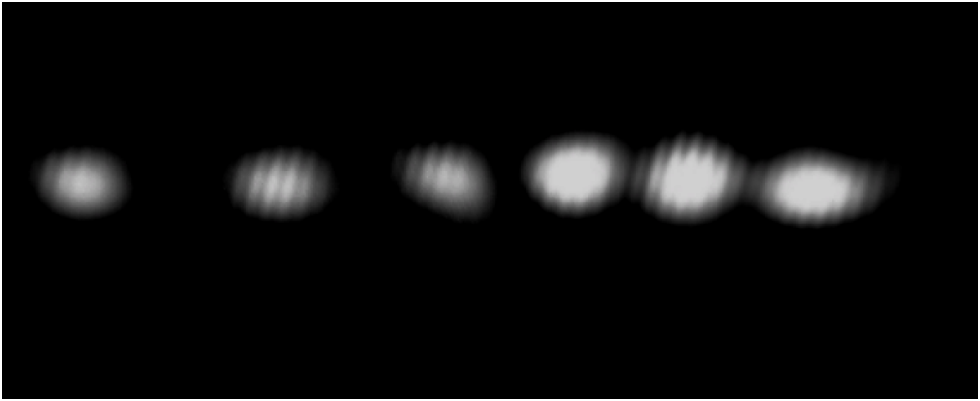}
\end{center}
\caption{\label{beads} An image of a string of 100\,nm diameter beads trapped in
the standing wave trap. A single bead is trapped by continually blocking and
unblocking one of the trapping beams until only one sphere is trapped. }
\end{figure}

Silica (SiO$_{2}$) nanospheres, manufactured by Microspheres-nanospheres and
Bangs Laboratories, are introduced into the trapping region at atmospheric
pressure via an ultrasonic nebulizer (Omron NE-U22). These spheres range in
radius from 26\,nm to 510\,nm and are suspended in methanol. The nanosphere
solution is sonicated using an ultrasonic bath for at least an hour before
trapping to prevent clumping. Once introduced into the trapping region the
methanol surrounding the spheres rapidly evaporates and the spheres are trapped
over many fringes of the standing wave, as shown in Fig.~\ref{beads}. As our
imaging system does not have single fringe resolution we cannot determine if
more than one sphere is trapped in a single fringe by this method. However, this
information can be inferred from the relative intensity of the light scattered
from the trapped spheres and also by the reduced stability of the particles in
the trap when more than one particle is trapped. To reduce the number of trapped
particles the 
trapping light is briefly blocked and unblocked. This is repeated until a single
sphere is visible in the trap. At this pressure, where there is a strong damping
force from air the sphere can be held in the trap indefinitely.

\begin{figure}[htb]
\begin{center}
\includegraphics[height=2.0in]{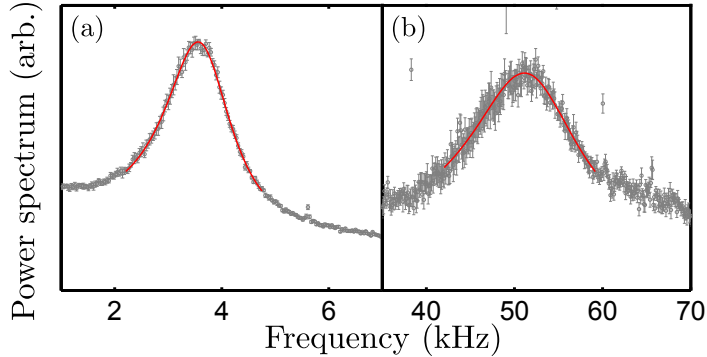}
\end{center}
\caption{(Colour online) Power spectra at 5\,mbar calculated from a measurement
of the position of a trapped 200.1\,nm diameter nanosphere as a function of
time, using a QPD. (a) The transverse frequency, and (b) an axial frequency.
Outlier points are due to electronic noise. Red lines show fitted Gaussian
functions, from which the trap frequencies are extracted.}
\label{fft}
\end{figure}

To measure the trap frequency the air is pumped from the system, and at this
point no more spheres enter the trap, as without air-damping their velocity is
too high. The air pressure in the trap is reduced to 5\,mbar, so that clear trap
frequencies can be obtained from the power spectrum of the position fluctuations
of the trapped sphere, as recorded on the QPD. Example power spectra are shown
in Fig.~\ref{fft}. Above 5\,mbar the damping of the motion in the trap due to
air broadens the peak in the power spectrum so that finding an accurate trap
frequency is difficult. Below pressures of 5\,mbar the spheres become unstable
in the trap and escape. This is most likely due to radiometric forces which have
been compensated for in other experiments using feedback techniques
\cite{Ashkin,Raizen}.  At 5\,mbar the damping rate due to gas collisions is
significantly less than our lowest measured trap frequencies, and thus the
measured frequency at this pressure is a good approximation to the bare trap
frequency which would be measured in vacuum without damping.

The angular axial trap frequency for a small polarizable particle in a standing
wave is $\omega_a = 2\pi f_a = \sqrt{\frac{4 \alpha k^2 I_0}{m\epsilon_0 c}}$,
where the polarizability of a sphere of refractive index $n$ is $\alpha = 4 \pi
\epsilon_0 r^3 \frac{n^2-1}{n^2+2}$. The maximum intensity in the radial center
of each equal intensity beam is given by $I_0$, and $k$ is the magnitude of the
wavevector of each beam. The sphere has mass $m= 4/3 \pi\rho r^3$, radius $r$
and density $\rho \simeq 2000$\,kg\,m$^{-3}$. The transverse trap frequency is
given by $\omega_t=\sqrt{\frac{8\alpha I_0}{ m\epsilon_0 c w^2}}$, where $w$ is
the focused spot size (radius) of the two counter-propagating beams. From these
expressions the ratio of the trap frequencies is given by $\omega_a/\omega_t=k
w/\sqrt{2}$.

\begin{figure}[htb]
\begin{center}
\includegraphics[height=3.5in]{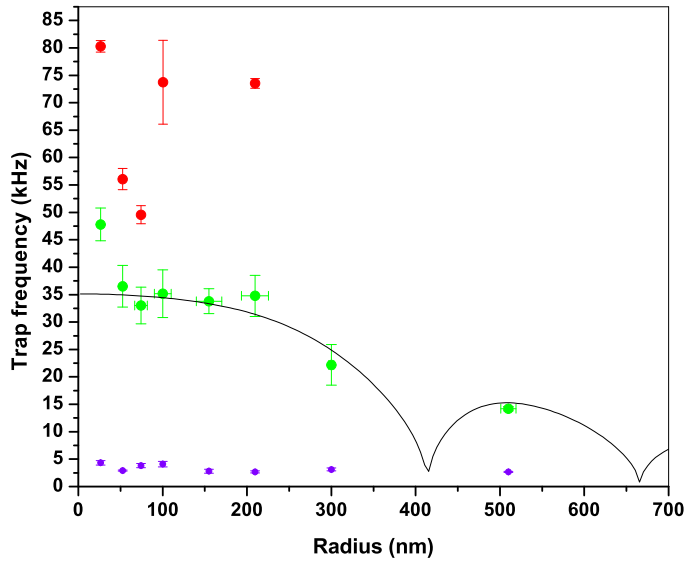}
\end{center}
\caption{(Colour online)  Measured trap frequencies as a function of sphere
radius. Points plotted in green are the axial trap frequency, blue are the
transverse trap frequency and the red data points are the higher axial trap
frequencies which are believed to be due to optical binding. The solid black
line is a theoretical curve derived from a numerical calculation \cite{Barton}.}
\label{exp_trap_freq}
\end{figure}

The trap frequencies in each axis are determined by fitting measured position
fluctuation power spectra using $\frac{2 k_B
T}{m}\frac{\Gamma_0}{(\omega_a^2-\omega^2)^2+\omega^2 \Gamma_0^2}$, where $k_B$
is Boltzmann's constant and $\Gamma_0$ is the damping rate. The fit to the data
is shown in Fig.~\ref{fft} with $\Gamma_0/2\pi \simeq 2.4$\,kHz for the axial
trap frequency. Several sets of data were taken for different spheres of the
same nominal radius and the measured axial and transverse trap frequencies for
each size sphere are shown in Fig.~\ref{exp_trap_freq}. The derived trap
frequencies for each sphere radius are the average over different experiments at
each radius, and the errors are the standard errors in the mean. The uncertainty
in the sphere radius is taken from the information supplied by the manufacturer.
Two axial (red and green data in Fig.~\ref{exp_trap_freq}) frequencies and one
transverse frequency (blue data points) were measured. When a particle is
tightly trapped by the optical field only one axial 
frequency is expected from a single sphere in a standing wave.  The lower axial
frequency (in green in figure~\ref{exp_trap_freq}) is always observed in the
data and this is taken as the true axial frequency. The higher frequency, which
is often present in the data, may be due to the trapping of two spheres in a
single anti-node, with the higher frequency occurring due to optical binding,
which requires further study \cite{Zamenek}. The higher axial frequency also
changes rapidly with sphere size, indicating that it is not the true axial trap
frequency, which should be almost constant for the small spheres. The presence
of a single axial frequency is, we believe indicitave of having trapped a single
sphere.

Although we don't know the radial dimensions of the beam within the trap we can
estimate this value from the ratio of the axial to transverse trap frequencies
for small spheres. The spot size from this ratio is
$w=\frac{\sqrt{2}}{k}\frac{\omega_a}{\omega_t}$ and for
$\frac{\omega_a}{\omega_t}$=9.8 this gives a spot size of $w= 2.3\,\mu$m.  Since
the trap frequency with two overlapping beams of size 2.3\,$\mu$m would be equal
to 207\,kHz with a power in each beam of 150\,mW, and we only measure a maximum
axial trap frequency of approximately 40 kHz, we conclude that the particles are
trapped in a standing wave formed where the waist of one beam is much larger. If
one spot size is 2.3\,$\mu$m the other would have to be 15\,$\mu$m. A plot of
the calculated axial trapping frequency, found by calculating Maxwell's stress
tensor (\cite{Barton,Chang}), is also shown in Fig.~\ref{exp_trap_freq}. Like
the experimental data the trapping frequency is constant for small spheres and
decreases to approximately zero when 
the particle size is comparable to the size of the interference pattern produced
by the standing wave. At larger radii the force on the particle changes sign and
a stable trap is formed in a node of the standing wave, as shown for the
particle of radius 510\,nm. Our measurements confirm that for particle radii
less that 200\,nm the simple dipole model for the nanospheres is adequate for
modelling the cooling and dynamics of the nanospheres in an optical cavity
utilising 1064\,nm radiation. 

Our experiments have shown that optical traps without feedback are currently
limited to operation at pressures down to a few millibar for all particles that
we have measured. In addition this limiting pressure did not change by reducing
the intensity by 50\%.  This radiometric force is due to localized heating of
nanosphere and the subsequent heating of the surrounding air. At low pressures,
when the mean free path is comparable to the size of the nanosphere, the radiometric force competes with and
eventually dominates the dipole force which traps the particle. While feedback
techniques have been successful \cite{Raizen}, decreasing the absorption of the
nanospheres is another route to minimising radiometric effects. This is feasible since
all the spheres we have used in this study are not made of optical quality glass
but from colloidally grown nanospheres. Finally, we have also successfully
trapped silica spheres in an ion trap at pressures of $10^{-6}$\,mbar which could be
used to load an optical trap formed by a cavity at lower pressures where 
radiometric forces are not significant.

\section*{Conclusions}
We have described a study of the dynamics and noise spectra of self-trapped
levitated optomechanical
systems. We have been able to show, by combining experimental measurements and
theoretical
calculations that strong light-matter coupling is attainable over a wide range
of particle sizes, and that these can be trapped.
The interdependence of the mechanical and optical mode frequencies, unique to
self-trapped levitated
systems provides a complex and interesting side-band structure, including
multi-mode mixing and bistabilities which we aim to explore experimentally.
These conclusions are supported by measurements of trap frequency made in an
optical standing trap where we have demonstrated a protocol for loading a single
nanosphere in a single antinode. 

{\em Acknowledgements}: We acknowledge support for the UK's Engineering and Physical Sciences Research Council.

\end{document}